\documentclass[conference]{IEEEtran}
\IEEEoverridecommandlockouts
\usepackage{cite}
\usepackage{amsmath,amssymb,amsfonts}
\usepackage{algorithmic}
\usepackage{graphicx}
\usepackage{textcomp}
\usepackage{xcolor}
\def\BibTeX{{\rm B\kern-.05em{\sc i\kern-.025em b}\kern-.08em
    T\kern-.1667em\lower.7ex\hbox{E}\kern-.125emX}}
\begin{document}

\title{A New Era of Mobility: Exploring Digital Twin Applications in Autonomous Vehicular Systems}

\author{\IEEEauthorblockN{S M Mostaq Hossain}
\IEEEauthorblockA{\textit{Dept. of Computer Science} \\
\textit{Tennessee Tech University}\\
Tennessee, USA \\
shossain42@tntech.edu}
\and
\IEEEauthorblockN{Sohag Kumar Saha}
\IEEEauthorblockA{\textit{Dept. of Electrical \&} \\
\textit{Computer Engineering} \\
\textit{Tennessee Tech University}\\
Tennessee, USA \\
ssaha42@tntech.edu}
\and
\IEEEauthorblockN{Shampa Banik}
\IEEEauthorblockA{\textit{Dept. of Computer Science} \\
\textit{Tennessee Tech University}\\
Tennessee, USA \\
sbanik42@tntech.edu}
\and
\IEEEauthorblockN{Trapa Banik}
\IEEEauthorblockA{\textit{Dept. of Electrical \&} \\ 
\textit{Computer Engineering} \\
\textit{Tennessee Tech University}\\
Tennessee, USA \\
tbanik42@tntech.edu}

}

\maketitle

\begin{abstract}

Digital Twins (DTs) are virtual representations of physical objects or processes that can collect information from the real environment to represent, validate, and replicate the physical twin's present and future behavior. The DTs are becoming increasingly prevalent in a variety of fields, including manufacturing, automobiles, medicine, smart cities, and other related areas. In this paper, we presented a systematic reviews on DTs in the autonomous vehicular industry. We addressed DTs and their essential characteristics, emphasized on accurate data collection, real-time analytics, and efficient simulation capabilities, while highlighting their role in enhancing performance and reliability. Next, we explored the technical challenges and central technologies of DTs. We illustrated the comparison analysis of different methodologies that have been used for autonomous vehicles in smart cities. Finally, we addressed the application challenges and limitations of DTs in the autonomous vehicular industry.

\end{abstract}


\section{Introduction}
Over the past decade, autonomous driving (AD) has grown fast, transforming the transportation system in terms of safety and efficiency \cite{b1}. As AVs proliferate, safety and dependability are paramount in AV system development. The latest research implies that AVs could considerably improve vehicle safety \cite{b2}, but this won't be achievable until a fleet of AVs has tested billions of kilometers in all weather situations. Running a fleet of AVs and development infrastructures that use physical testing data would take decades and tens of billions of dollars to meet AV \cite{b3} safety targets. 

Digital Twin (DT) technology has various uses, from real-time remote monitoring and control in industry to risk assessment in transportation to smart scheduling in smart cities, therefore it has received a lot of attention recently. According to Z. Hu et al. \cite{b4}, figure 1 shows the major DT development milestones. In a high-fidelity virtual environment, simulation-based digital twins can speed up AV verification and save development expenses \cite{b5}. Software must change for digital twins. Digital twins can simulate various environmental and traffic situations, avoiding the need for extensive physical testing. Virtual, controllable testing of autonomous vehicles could save several orders of magnitude in development time and cost. We may have to wait months for significant snow to test AVs on the road. A digital twin can build a road, simulate a major snowstorm, and generate a lot of high-quality testing data \cite{b6}. Our initial deployment of a digital twin system that combines physical and digital twin testing has been successful, but its flexibility allows for further improvement \cite{b7}.

Vehicle dynamic simulators, including cruise control system simulators, have been widely used in the automobile industry for testing \cite{b8}. Aerospace has long used simulation. Testing autonomous vehicles in virtual and controllable environments could save development costs and time by orders of magnitude.

\begin{figure*}
  \includegraphics[width=\textwidth]{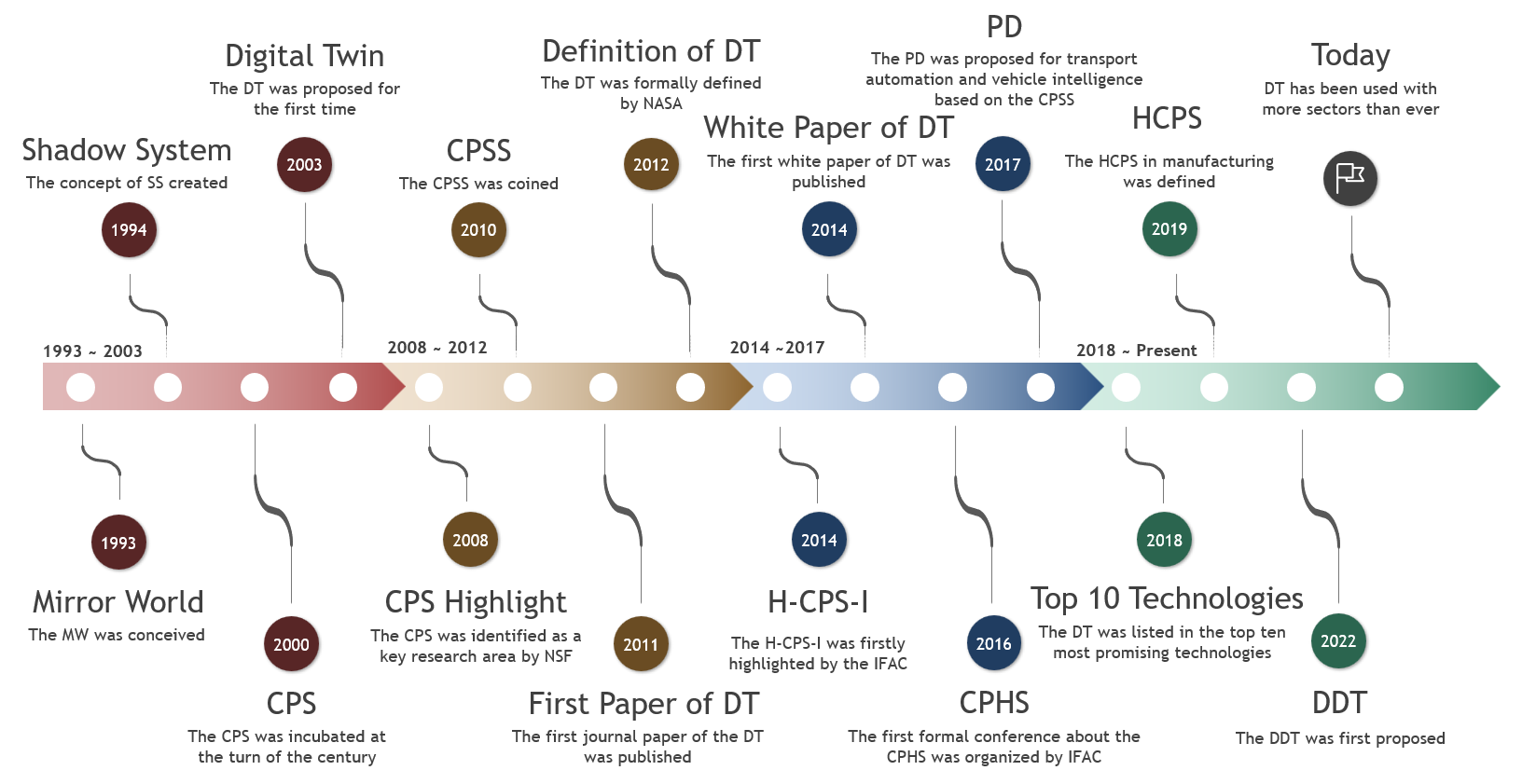}
  \caption{History of digital twin technology}
\end{figure*}

In the development of AD software , simulators have been used extensively to test and evaluate the decision-making module and path-planning module under various scenarios by providing perception data (such as the position and moving states of the ego vehicle and other traffic participants) \cite{b9}. This strategy is easy and scalable, however it does not represent the reality well, which causes problems. Simulation tests cannot compare to physical AV software pipeline tests. This pipeline includes sensing, localisation, decision-making, path planning, and vehicle control. Physical examinations. Physical elements, such as weather and lighting, also prevent investigations.

First, these simulators use virtual town maps instead of real-world road testing \cite{b10}. Not simulated are road conditions. A digital twin map is needed to evaluate AD functions like exit or entry highway ramps that depend on road geometry and traffic legislation. Simulators' car and pedestrian animations are pre-programmed. Simulations cannot replicate real traffic's intricate interactions. Junctions and aggressive driving are unjudgable. AV software testing suffers from low-fidelity sensor data. Depth mapping and ray casts replicate lidar sensors but don't account for reflection and diffusion. Simulations differ. Unlike AV software development, automobile hardware development accelerates with "digital twin" physical simulation tools like MATLAB and Modelica.

Figure 2 shows how our two-tiered framework creates a linked vehicle digital twin system \cite{b11}. Virtual is above actual. This system's communication module is vital. Cellular data transport powers this experiment. The physical layer of the digital twin framework can represent all physical entities and their interactions, such as automobiles and their parts, drivers and passengers, road infrastructure, weather, other road users, etc., defined on a global coordinate system and developing over time. Sensors and actuators. The sensors may detect and aggregate vehicle speed, driver gaze, and traffic light status at various resolutions. The communication module analyzes data online. The item or process, sensors, actuators, and computing resources are shown in Figure 2 \cite{b13}. The digital world comprises databases, data processing infrastructures, machine learning, and the digital twin. Wi-Fi and Bluetooth protocols and interfaces link them. Architecture monitoring requires visualization.

The key contributions of this work are as follows: 
\begin{itemize}
    \item We investigated the overview and recent researchers on the DTs concept specifically for autonomous vehicular systems.
    
    \item The state-of-the-art research methodologies are discussed based on current literature's of DTs.
    
    \item The comparison of different methods, their role in technical development and limitations are stated.
\end{itemize}

The remaining sections of this paper is organized in the following manner: Section II provides a technological overview, Section III presents a brief literature review, Section IV outlines the methodologies used, and Section V presents a comparative analysis. The remaining sections such as: Section VI and Section VII, consists of the Discussion and Conclusion, respectively.

\begin{figure*}
  \includegraphics[width=\textwidth]{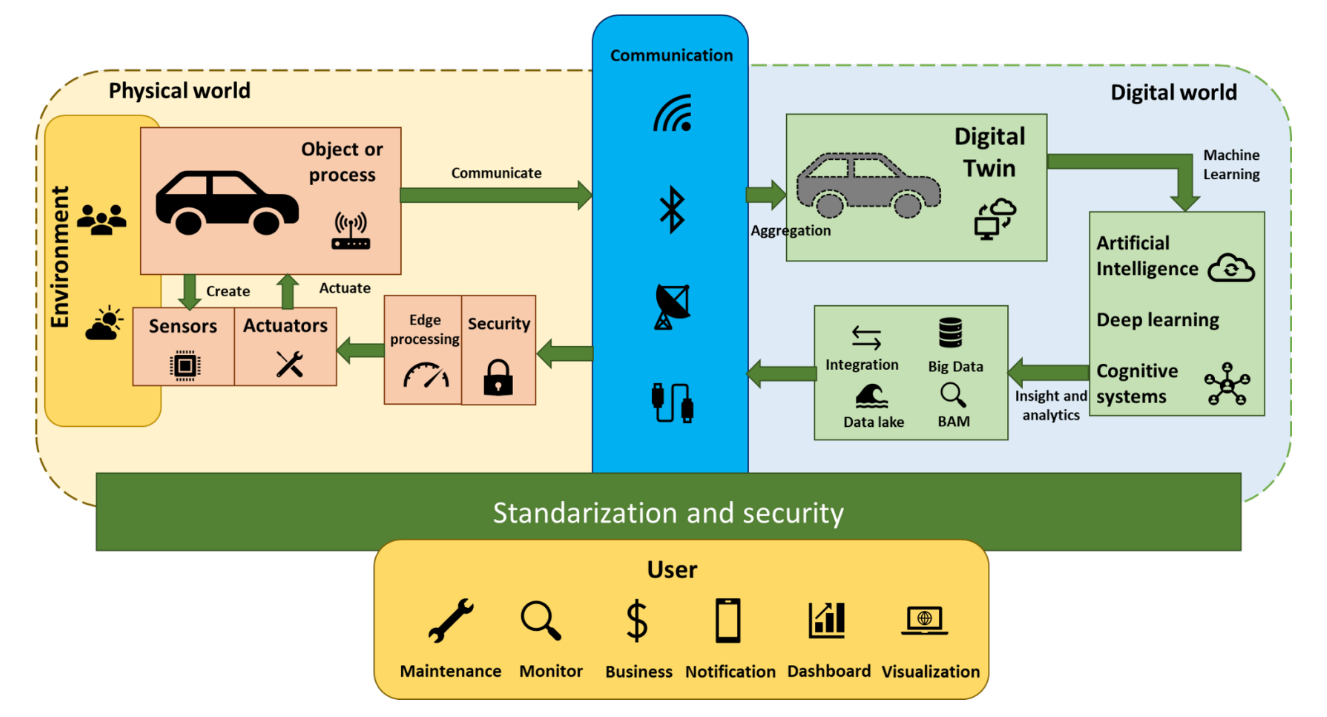}
  \caption{General framework of the digital twin system for connected vehicles \cite{b11}}
\end{figure*}

\section{An overview of digital twin technology for smart vehicles}
Digital twins are used in the automotive industry to create digital copies of vehicles. Data on car use and performance enables for more personalized service and maintenance. Digital copies can be model copies or networked systems \cite{b14}. Engineers investigate AI before sending a car to the assembly line. Simulation models can predict breakdowns and wear. Instead of road testing and maintenance, autonomous vehicle digital twins could save unforeseen costs. Digital twin technology may imitate and improve many aspects of a smart electric vehicle's system, which has far-reaching effects \cite{b15}. Digital twin vehicle modeling requires understanding the digital twin environment.

Automakers employ digital twins to make digital duplicates of automobiles. Car performance data allows for more targeted service and maintenance. Model copies or networked digital copies \cite{b14}. Engineers study AI before assembling an automobile. Simulations forecast breakdowns and wear. Autonomous vehicle digital twins could eliminate road testing and maintenance. Digital twin technology can replicate and improve a smart electric vehicle's system. Digital twin vehicle modeling requires digital twin environment knowledge.


\section{Literature Review}
In this section, we've presented a brief overview of the DT concept for the autonomous vehicular industry. The following discussion has included the literature overview regarding our goal.

In the paper, \cite{b16} B. Yu, et al. shared real-world experiences of the digital twins, a practical method for developing autonomous driving (AD) systems that creates a complete, accurate, and reliable model of the physical environment to reduce the need for physical testing. Their main contributions are:
\begin{itemize}
    \item they have identified the limitations of conventional approaches to AD simulation and show how digital twins can be used to overcome them.
    \item To begin, they synthesized their practical development experience into three overarching concepts for the AD digital twin system's design.
    \item at the end, they described the AD digital twin system's structure and its components, including how real-world mapping data is collected, sensor data is mimicked, and traffic actors are synthesized.
\end{itemize}

The followings are the paper's \cite{b17} unique contributions in comparison to previous recent research on the validation of coordination strategies:
\begin{itemize}
    \item At the end, they described the AD digital twin system's structure and its components, including how real-world mapping data is collected, sensor data is mimicked, and traffic actors are synthesized.
    \item Under the digital twin paradigm, a working prototype has been created. Time lag and precision of localization are just two of the worrying metrics tracked.
    \item The system includes HMI devices. In 3D, the Hololens controls automobiles. First-person perspective driving simulators allow drivers full control.
\end{itemize}

The study \cite{b18} examined digital twin technology's origins and deployment phases. This research highlights digital twin technologies like predictive mobility, autonomous motion control, driver assistance systems, vehicle health management systems, battery management systems, intelligent charging, vehicle power electronic converters, and electric power drive systems. Barriers to adoption and important supporting technologies are also identified, which will aid future eco-friendly and sustainable transportation endeavors.

The contribution of the paper \cite{b19} are follows:
\begin{itemize}
    \item In densely populated locations with considerable traffic, automobiles are increasingly being used as services due to commercial expansions. This study suggested using DT to enable CaaS.
    \item The suggested design includes the city, a middleware to connect all entities, DTs models that run over the middleware, and applications like car-sharing.
    \item The case study showed how the proposed concept might be implemented and highlighted several key areas for development in future efforts.
\end{itemize}

The paper \cite{b20} presented a Petri-net-based DT to simulate the electric car development process from start to finish. Real-time data sharing between the physical system and its digital shadow can help make better, faster decisions. The two systems can directly calculate and implement actions to contain these facts, calculate new time plans, and inform the user of optimistic, most likely, and pessimistic scenarios for task delays. Our research will improve search algorithms for non-computable solutions and optimize physical-digital subsystem communication and interaction.

This study proposes a V2C-communicating linked car digital twin structure \cite{b21}. The vehicle's driver-vehicle interface (DVI) displays the cloud server's advisory speed, letting the driver control the vehicle. ADAS uses the digital twin paradigm. The suggested digital twin structure is tested in real-world traffic on three passenger vehicles in a cooperative ramp merging case study.

In paper \cite{b22}, the author's plan is to create an unsupervised prognosis and control platform tailored to electric propulsion drive systems (EPDS) performance estimation as a final product. A number of subsidiary tasks and goals must be formulated in order to accomplish this primary aim.

\begin{itemize}
    \item Construct the DT of the energy system by creating physical models of its constituent parts (motors, generators, gearboxes, bearings, etc.) and their corresponding reduced models (testbed).
    \item Create a working prototype of the Virtual Sensors idea, which is built on the existing DT concept.
    \item Make a system based on artificial intelligence that lets the virtual sensors be used to control EPDS.
    \item Demonstrate the understanding of these ideas and how they apply to achieve the aforementioned end goal using the autonomous vehicles as a case study.
\end{itemize}

\begin{table*}[]
\caption{Comparison analysis of different digital twin systems for autonomous vehicle system}
\label{tab:my-table}
\begin{tabular}{|l|l|l|l|l|l|}
\hline
\textbf{Sl.} & \textbf{Methodology Used}& \textbf{Role of DT Technology}& \textbf{Evaluation}& \textbf{Future Aspects}& \textbf{Year {[}Ref.{]}} \\ 

\hline

1 & \begin{tabular}[c]{@{}l@{}}DT-GM: SUMO's innovative \\microscopic simulation-based \\Geneva highway digital twin.\end{tabular}
& \begin{tabular}[c]{@{}l@{}}DT-GM's development process\\ is explained, highlighting \\SUMO's calibration features \\that allow ongoing calibration \\of running simulation scenarios.
\end{tabular}                                                
& \begin{tabular}[c]{@{}l@{}}Experimental findings show\\that DT-GM accurately\\represents traffic. ODPMS's \\current development enables\\ DT-GM's low-latency \\motorway response.\end{tabular}
& \begin{tabular}[c]{@{}l@{}}Usefulness of powerful \\self-regulating adaptive \\controllers to optimize\\variable speed restrictions\\ in safety-critical\\ decisions using DT.
\end{tabular}   & 2023 \cite{b38} \\

\hline

2 & \begin{tabular}[c]{@{}l@{}}Examined the IoDT's \\architecture, communication \\modes, major aspects, \\supporting technologies, \\and recent prototypes.\end{tabular}
& \begin{tabular}[c]{@{}l@{}}We explore IoDT security \\and privacy challenges \\from seven perspectives—data,\\ authentication, communication, \\privacy, trust, monetization, \\and cyber-physical—and the \\main obstacles to tackling them.\end{tabular}                                                
& \begin{tabular}[c]{@{}l@{}}Comprehensive understanding \\of IoDT working principles, \\including its general architecture,\\ key characteristics, security or \\privacy threats, and existing or \\potential countermeasures.\end{tabular}
& \begin{tabular}[c]{@{}l@{}} Cloud-Edge-End Orchestrated,\\ Space-Air-Ground Integrated,\\ Interoperable and Regulatory,\\ Explainable AI-Empowered,\\  Information Bottleneck Based, \\ and Privacy-Aware of IoDT\end{tabular}   & 2023 \cite{b39} \\

\hline

3 & \begin{tabular}[c]{@{}l@{}}Developed a digital twin a \\ system using a game engine \\ for 3D modeling, picture \\ rendering, and physical \\ simulation. Structural, \\ physical, and logical twins \\ are introduced.\end{tabular}                                                                                                           & \begin{tabular}[c]{@{}l@{}}Three components implement \\ digital twin properties: The \\ 3D digital twin map is \\ structural, sensor models are \\ physical, and the traffic \\ controller is logical.\end{tabular}                                                & \begin{tabular}[c]{@{}l@{}}The digital twin paradigm creates \\ an entire, comprehensive, exact, \\ and trustworthy representation of \\ the physical environment, \\ allowing for quick, low-cost \\ development iterations.\end{tabular}
& \begin{tabular}[c]{@{}l@{}}Digital twin cost and \\ efficiency can be \\ improved by optimizing \\ the computing system.\end{tabular}   & 2022 \cite{b16} \\ 

\hline

4                & \begin{tabular}[c]{@{}l@{}}The archetypal infrastructure \\ uses sand table, clone, and \\ cloud. A driving simulator, \\ Hololens, and a screen display \\ improve system engagement. \\ A platoon based test proves \\ the system can test many \\ cars simultaneously.\end{tabular}                                                         & \begin{tabular}[c]{@{}l@{}}This research advises \\ employing digital twins \\ instead of real automobiles \\ for multi-vehicle studies. \\ The system is also modeled.\end{tabular}                                                                                                                       & \begin{tabular}[c]{@{}l@{}}Virtual speed fluctuations are \\ smoother, and speed profiles \\ between the same type of \\ vehicles are comparable, but \\ those between different types \\ vary greatly. To follow a physical \\ vehicle, V2's spacing variations \\ are unsteady, like small ones.\end{tabular}                                                                                  & \begin{tabular}[c]{@{}l@{}}1) Impact of different \\ access modes on \\ experiment results \\ will be analyzed. \\ 2) More dynamics \\ models of cloud \\ vehicles will be provided\end{tabular}                                                                                            & 2022 \cite{b17}                        \\ \hline
5                & \begin{tabular}[c]{@{}l@{}}This article presents an \\ eclectic overview of the \\ smart car system by \\ discussing each \\ component in depth.\end{tabular}                                                                                                                                                                                  & \begin{tabular}[c]{@{}l@{}}The study includes predictive \\ mobility, autonomous motion \\ control, driver aid systems, \\ vehicle health management \\ systems, battery management \\ systems, and electric power \\ drive systems.\end{tabular}                                                          & \begin{tabular}[c]{@{}l@{}}1) Generating historical collision \\ time and speed data from case \\ scenarios. 2) Use real-time vehicle \\ data and a machine-learning model \\ to implement telematics-based \\ ADAS. 3) Predict driving safety \\ using historical data, sensor \\ fusion, and privacy policy.\end{tabular}                                                                      & \begin{tabular}[c]{@{}l@{}}AI can help forecast \\ future EV performance \\ metrics. Cloud-based \\ digital twin technology \\ reduces storage on \\ mobile systems like \\ vehicles by decentralizing \\ data storage.\end{tabular}                                                        & 2021 \cite{b18}                        \\ \hline
6                & \begin{tabular}[c]{@{}l@{}}The AutomationML is a \\ well-defined modeling tool \\ and is widely adopted by \\ companies. It allows \\ reusability and \\ interoperability between \\ different applications and \\ languages.\end{tabular}                                                                                                     & \begin{tabular}[c]{@{}l@{}}A concept of enabling and \\ supporting CaaS by Digital \\ Twin was proposed and a \\ use case has been implemented \\ to demonstrate how it can be \\ applied in a real scenario.\end{tabular}                                                                                 & \begin{tabular}[c]{@{}l@{}}Case study illustrates how Digital \\ Twin supports Car-as-a-Service. \\ All entities have virtual \\ representations with middleware. \\ This system tracks automobile, \\ user, and user-car data to improve \\ future user experiences.\end{tabular}                                                                                                               & \begin{tabular}[c]{@{}l@{}}Before implementing \\ this strategy, consider \\ safety and security. \\ Blockchain-based \\ technologies may \\ help with this. \\ Wearables can help \\ build a reactive app.\end{tabular}                                                                    & 2021 \cite{b19}                        \\ \hline
7                & \begin{tabular}[c]{@{}l@{}}To handle delays, Petri nets \\ are used to simulate \\ development tasks and \\ dependencies. The model is \\ connected to and interacts \\ with the physical system. \\ Alternative strategies to \\ overcome delays are \\ researched, and the ideal \\ option is computed, tested, \\ and applied.\end{tabular} & \begin{tabular}[c]{@{}l@{}}The Technical University of \\ Crete Eco Racing team \\ employed DTs to design, \\ manufacture, and assemble \\ a single-seat urban prototype \\ vehicle (TUCER). DT is not \\ a model of the vehicle but is \\ used to monitor and organize \\ development tasks.\end{tabular} & \begin{tabular}[c]{@{}l@{}}Real-time interaction between a \\ physical system and its digital \\ shadow allows for faster, more \\ accurate assessments. In case \\ of job delays, the two systems' \\ interface calculates and \\ implements actions to contain \\ them, computes new time plans, \\ and tells the user optimistic, \\ feasible, and pessimistic \\ possibilities.\end{tabular} & \begin{tabular}[c]{@{}l@{}}Optimizing communication \\ and interaction between \\ physical and digital \\ subsystems, adding \\ cognition to the Digital \\ twin, and using enhanced \\ search methods for \\ circumstances where no \\ practical answer can be \\ determined.\end{tabular} & 2021 \cite{b20}                        \\ \hline
8                & \begin{tabular}[c]{@{}l@{}}Vehicle to Cloud-based \\ advanced driver \\ assistance systems.\end{tabular}                                                                                                                                                                                                                                       & \begin{tabular}[c]{@{}l@{}}A digital twin framework is \\ proposed for connected \\ vehicles, which consists of a \\ physical layer and cyber layer \\ with various modules. As a \\ paradigm of this framework, \\ an advisory speed-based \\ ADAS is presented using \\ V2C communication\end{tabular}   & \begin{tabular}[c]{@{}l@{}}Three-passenger automobiles \\ are used to demonstrate the \\ digital twin framework's \\ usefulness in real-world traffic. \\ The digital twin can improve \\ transportation mobility and \\ environmental sustainability \\ with tolerable communication \\ delays and packet losses.\end{tabular}                                                                  & \begin{tabular}[c]{@{}l@{}}This study will evaluate \\ the digital twin concept \\ in mixed traffic, where \\ not all vehicles have \\ V2C connections. New \\ service modules must \\ be built and deployed in \\ this digital twin topology \\ to use V2C connectivity.\end{tabular}      & 2020 \cite{b21}                        \\ \hline
9                & \begin{tabular}[c]{@{}l@{}}To create an unsupervised \\ prognosis and control platform \\ tailored to electric propulsion \\ drive systems (EPDS) \\ performance estimation.\end{tabular}                                                                                                                                                      & \begin{tabular}[c]{@{}l@{}}Construct the DT of the energy \\ system by creating physical \\ models of its constituent parts \\ (motors, generator, gearboxes, \\ bearings, etc.) and their \\ corresponding reduced models.\end{tabular}                                                                   & \begin{tabular}[c]{@{}l@{}}Modeling can yield physical \\ device models (ex. MATLAB). \\ Physical models are reduced \\ in order. Parallel, real-time \\ reduced component models \\ can DT electric propulsion \\ drive systems.\end{tabular}                                                                                                                                                   & \begin{tabular}[c]{@{}l@{}}Digital twins can act as \\ virtual sensors or contain \\ virtual sensors. Combining \\ real and virtual sensor data \\ with machine learning can \\ diagnose electric energy \\ system devices.\end{tabular}                                                    & 2019 \cite{b22}                        \\ \hline
\end{tabular}
\end{table*}

\section{Methodologies of the selected pieces of literature}

There have been several methodologies have been discussed of the selected pieces of literature. Among them are discussed below.
\subsection{Three Principles architecture}
The authors have created a digital twin system using a game engine, which includes graphics and physics engines for 3D modeling, image rendering, and physical simulation, allowing for the representation of structural, physical, and behavioral information in a virtual world \cite{b16}. In order to implement the digital twin properties, three additional building blocks are added on top of the game engine: It can be seen that 1) the traffic controller is the logical twin, 2) sensor models are the physical twin, and 3) the 3D digital twin map is the structural twin. 
\subsection{Multi-vehicle Experiment Platform}
Tsinghua is developing a CAV-focused digital twin system. Even without real cars, this technology will allow multi-vehicle tests. This study \cite{b18} advises using real and virtual autos to get the desired outcome. A sand table testbed helps small cars run smoothly. A game engine creates cyberspace through full-element modeling. Cyberspace can show the sand table's real-time state. This research proposes a cloud vehicle to replace smaller autos.

\subsection{Car-as-a-Service Concept with DT}
Pana offers Car-as-a-Service (CaaS) employing sensors, actuators, and radio devices \cite{b23}. Smart cities should use linked automobiles for passenger service, the authors say. GNSS, high-precision distance estimation, radio connectivity, environmental sensors for measuring temperature, humidity, pressure, etc., motion sensors like an Inertial Measurement Unit (IMU) for measuring traffic flow, vehicle heading and roll, and road quality, a centralized cloud processing unit, and social network analytics are the backbone of CaaS \cite{b19}. CaaS may entail carpooling. Users rent and share cars. This service cuts traffic and saves occasional drivers. Archer \cite{b24} cites studies showing that five to fifteen privately owned cars are replaced for every shared car added to the fleet, assuming car-sharing programs reduce car ownership. Ferrero \cite{b25} investigates car-sharing. Technical and modeling studies have examined car-sharing businesses. This service is hard to sell. European, US, Japanese, Chinese, and Australian car-sharing schemes are widespread. Mattia \cite{b26} expects 12 million users by 2021. User settings \cite{b27} and driver monitoring will increase comfort and safety in this car-sharing concept. \cite{b28} found that Drive Monitoring Assistance System (DMAS) must monitor drivers' attention levels for safe driving. Distraction and fatigue cause most traffic accidents.

\subsection{Petri nets and variations}
Modeling, simulating, and tracking development Arc-extended timed Petri nets will be used \cite{b20}. Formally, an Ordinary Petri Net (OPN) is a five-tuple with these elements: PN requires a finite number of locations (P= p1, p2,..., pnp), transitions (T= t1, t2,..., tnt), vertices (V), and an empty set (PT=) as their intersection. In the PN token distribution, m0 is the initial token distribution, I and O are input and output functions (PNs initial marking). Locations in a Petri net network represent resources, governing conditions, transitions, and arcs \cite{b29}.

T-timed PNs implement transition time delays. They behave like Ordinary PNs but can predict event durations, making them better for simulation. Time-delayed PNs, or T-timed PNs, are defined as TPN= "P, T, I, O, m0, D," where D is a function of positive real integers. Using arc extensions to activate or deactivate PN sections when certain criteria are satisfied is critical for control. Standard, inhibitor, and activator arcs (illustrated as dashed vectors; \cite{b30}, \cite{b31}) are used across the literature. Arc extensions boost the baseline model's simulation capacity by allowing more complex ideas with fewer node connections.

\subsection{Vehicle-to-Cloud  Based  Advanced  Driver Assistance  Systems}
Authors create a two-layer digital twin for linked cars \cite{b21}. Cyber tops physical. Communication links two system framework levels. Study communication is cellular. The physical layer of the digital twin system, specified on a world coordinate throughout time, may contain cars, components, drivers, passengers, roadway infrastructure, meteorology, other road users, etc. Sensors and actuators are key layers. Vehicle speed, driver look, and traffic signal status can be detected by the sensors. Cyberspace processes data from the communication module.

Through the communication module, cyberworld beings and processes become corporeal. ADAS-equipped people can drive connected cars. Cyberworld actuation guidance guides the automatic controller or human driver of linked cars to make cooperative or intelligent motions, enhancing safety, mobility, and sustainability. The digital twin's cyber domain computes this two-layer system. Physical items and processes have important digital copies. Physical world data is cleaned, integrated, and time-synchronized. Pre-processed data can be kept in the database for digital traceability or sent to the data mining \& knowledge discovery module for machine learning. Data mining and knowledge discovery create the physical world model. Vehicle, driving, and traffic simulators can be modeled. Data refreshes cyberworld knowledge. Modeling/simulation aids prediction and decision-making. For system performance, physical actuators think.

\section{Comparison Analysis}
The findings in Table 1 show the different methodologies have been mentioned on those selected papers. The third column defines the role of the DT technologies for the corresponding methods and their use cases. Then the fourth column includes the evaluation of those methods. The fifth column finally includes the future aspects of those pieces of literary works. 


\section{Discussion}
\subsection{Application Challenges and Limitations}
The difficulties that may develop, according to the authors, may vary depending on the scope and integration complexity of the application. Based on the research that was analyzed, there were five major problems with DT technology implementation that were found to be universal across all fields. These problems adequately wrap up the investigation and answer the supplementary question. SQ1 and contribute to answering the primary research question as it stands now:
\subsubsection{Data-related concerns (trust, privacy, cyber security, convergence and governance, acquisition and large-scale analysis)}
If a behavior cannot be reduced to a set of numbers, it becomes far more challenging for designers to replicate it. Examples include ecological stability \cite{b32}, socioeconomic inequality \cite{b33}, and political instability \cite{b32}. These societal and environmental innovations will focus on preliminary SRL stages, where the potential influence on selected stakeholders, the larger society, and the environment is more understood. In addition, this difficulty is associated with Table 1's levels 3 and 4, where the complexity of DT implementations is exacerbated by the need to enrich models in real time and with bidirectional flow of information.
\subsubsection{A deficiency in DT implementation standards, guidelines, and regulations}
Lack of standards and acknowledged interoperability, particularly in the manufacturing industry, are cited as reasons for the current restrictions on DT implementations by the authors of \cite{b34}. Adopting a widespread, concrete understanding of DTs and their importance requires articles that explain the benefits, define ideas and structures of DTs, and review the state of the art of the technology. In addition, researchers can influence lower levels of the TRL by focusing on this specific topic through surveys and literature reviews, thereby increasing the dissemination of fundamental principles and concepts.
\subsubsection{High implementation costs owing to more sensors and computing power}
DT implementations are costly, therefore their use is constrained by the availability of such resources, which is often lacking in impoverished nations \cite{b35}. Achieving level 3 on Table 1's maturity spectrum is difficult because of the rise in required sensors and the resulting increase in complexity in data connectivity and processing (where the digital model needs to be enriched with real-time information). Practitioners are hampered by this difficulty in their pursuit of greater TRLs, where pilot systems are proven and DTs are integrated into commercial designs and widespread deployments.
\subsubsection{AI and big data for long-term and large-scale data analysis}
Big data algorithms and internet of things technology are powerful allies that can provide significant support to successful implementations of DT \cite{b36}. This is because DT systems generate and analyze a significant amount of data, and these two technologies work hand in hand. In addition to this, the information that is flowing from the many levels of indicator systems creates a problem in the process of defining uniform rules and standards. This challenge aims to effectively target levels 4 and 5 of the maturity spectrum, and if it is successful, it may make it possible to enable a bidirectional flow of information, control of the physical world from the digital model, and even autonomous operations and asset maintenance.
\subsubsection{Challenges associated with communication networks}
The development of superior communication standards like 5G is essential. In \cite{b37}, the authors discuss the importance of enabling real-time data connectivity and operational efficiency for the DT, as well as other benefits of using 5G technology in smart cities, such as the ability to connect many more sensors and devices at high speeds with ubiquitous connectivity, improved reliability and redundancy, and ultra-low power consumption.
\section{Conclusion}
The digital twin technology opens up new opportunities for the development of sustainable electric vehicle technologies in terms of both cost-effectiveness and efficiency, beginning with the design phase and continuing through the operation. Because its sister technologies, such as the internet of things, decentralized wireless networking, and artificial intelligence, were not as far along in their development at the time as digital twin technology was being conceived, the car industry has only just begun to implement it. Nevertheless, a prime period for the development of digital twin technologies and other smart development approaches has now begun in the field of science, which has now entered this era. In addition, the next decade will mark a significant turning point in human history as a result of the extraordinary environmental difficulties that will be faced. As a result, the fundamental objective of the research community should be to foster sustainable technology and the enablers of such technologies. With this goal in mind, the purpose of this review is to demonstrate the implementation of digital twin technology in the automobile industry, with a focus on smart electric vehicles serving as the background for the discussion. In this article, the history of digital twin technology along with its development and the many stages of its deployment are discussed. This study focuses on the digital twin technologies that have been adapted for smart electric vehicle use cases. Some of these technologies include predictive mobility, autonomous motion control, driver assistance systems, vehicle as a service system, Petri net model discussion, and vehicle-to-cloud-based driver assistance systems.


\end{document}